\documentclass[aps,onecolumn,preprintnumbers,amsmath,amssymb,nofootinbib,superscriptaddress,notitlepage,11pt]{revtex4-1}

\usepackage{inputenc}
\usepackage{txfonts}
\usepackage{epsfig}
\usepackage{slashed}
\usepackage{color}
\usepackage{overpic}
\usepackage{subfigure}
\usepackage{hyperref}

\begin{document}

\title{Can we understand the decay width of the $T_{cc}^+$ state?}

\author{Xi-Zhe Ling}
\affiliation{School of Physics, Beihang University, Beijing 102206, China}

\author{Ming-Zhu Liu}
\email{zhengmz11@buaa.edu.cn}
\affiliation{School of Space and Environment, Beihang University, Beijing 102206, China}
\affiliation{School of Physics, Beihang University, Beijing 102206, China}

\author{Li-Sheng Geng}~\email{lisheng.geng@buaa.edu.cn}
\affiliation{School of Physics, Beihang University, Beijing 102206, China}
\affiliation{Beijing Key Laboratory of Advanced Nuclear Materials and Physics, Beihang University, Beijing, 102206, China}
\affiliation{School of Physics and Microelectronics, Zhengzhou University, Zhengzhou, Henan 450001, China}

\author{En Wang}~\email{wangen@zzu.edu.cn}
\affiliation{School of Physics and Microelectronics, Zhengzhou University, Zhengzhou, Henan 450001, China}

\author{Ju-Jun Xie}~\email{xiejujun@impcas.ac.cn}
\affiliation{Institute of Modern Physics, Chinese Academy of Sciences, Lanzhou 730000, China}
\affiliation{School of Nuclear Science and Technology, University of Chinese Academy of Sciences, Beijing 101408, China}
\affiliation{School of Physics and Microelectronics, Zhengzhou University, Zhengzhou, Henan 450001, China}
\affiliation{Lanzhou Center for Theoretical Physics, Key Laboratory of Theoretical Physics of Gansu Province, Lanzhou University, Lanzhou, Gansu 730000, China}

\date{\today}


\begin{abstract}
  \rule{0ex}{3ex}

  Inspired by the recent discovery of a doubly charmed  tetraquark state $T_{cc}^+$  by the LHCb Collaboration,  we employ the effective Lagrangian approach to investigate  the decay width of $T_{cc}^{+}\to D^{+} D^{0}\pi^{0}/D^{0} D^{0}\pi^{+}$ and $T_{cc}^{+}\to D^{0}D^{+}\gamma$ with the assumption that $T_{cc}^{+}$ is an isoscalar $DD^{\ast}$ molecule. We show that both the  $T_{cc}\to D D\pi$ and $T_{cc}\to DD\gamma$  modes contribute to  the decay width of $T_{cc}$, with the former being dominant. 
The resulting total decay width of about $\Gamma=63$ keV  is smaller than the experimental decay width obtained from the Breit-Wigner fit of the LHCb data, $\Gamma=410\pm 165\pm  43^{+18}_{-38}$ keV, while  close to the number obtained  from the alternative unitary analysis, $\Gamma=48\pm 2^{+0}_{-14}$ keV, which supports the molecular nature of $T_{cc}$.  
\end{abstract}

\maketitle

\section{Introduction}

Mesons made of a pair of quark and anti-quark and baryons made of three quarks can be well understood in the conventional quark model~\cite{Gell-Mann:1964ewy}.  Although Quantum ChromoDynamics (QCD)  allows for other quark configurations, such as tetraquark, pentaquark, and hexaquark states, their existence were not experimentally confirmed until the charged tetraquark state $Z_c(4430)$, with the minimum quark content $c\bar{c}u\bar{d}$, was discovered in 2007 by the Belle Collaboration~\cite{Belle:2007hrb}. In 2015, the LHCb Collaboration reported the first pentaquark states $P_{c}(4380)$ and $P_{c}(4450)$~\cite{LHCb:2015yax}, while the latter  was shown to be a superposition of  two states, $P_{c}(4440)$ and $P_{c}(4457)$~\cite{LHCb:2019kea}. In 2020, the LHCb Collaboration discovered the first fully heavy tetraquark state $X(6900)$~\cite{LHCb:2020bwg}.
It should be noted that all of these exotic states carry hidden charm number. The first open charm tetraquark states, $X_{0}(2866)$ and $X_{1}(2904)$, were only discovered in 2020 by the LHCb Collaboration~\cite{LHCb:2020bls}.

  Since forty years ago, a series of pioneer works  have already investigated the likely existence of $QQ\bar{q}\bar{q}$ tetraquark states in the quark model, which showed that the stability of  a $QQ\bar{q}\bar{q}$ tetraquark depends on the mass ratio of   $m_{Q}/m_{\bar{q}}$  ~\cite{Ader:1981db,Zouzou:1986qh,Lipkin:1986dw,Heller:1986bt,Carlson:1987hh,Silvestre-Brac:1993zem,Semay:1994ht}.
As the ratio is larger, such a  multiquark state is more stable.  However, due to the uncertainty of  $m_{c}/m_{\bar{q}}$,  whether the mass of  the $cc\bar{q}\bar{q}$ tetraquark state is above or below the $DD^{\ast}$ mass threshold is unsettled.  Later, meson exchange potentials were employed to study the likely existence of  $DD^{\ast}$ molecules~\cite{Manohar:1992nd,Pepin:1996id}, where due to the unknown parameters the existence of $DD^{\ast}$ bound states is also uncertain.  In 2002,  a 
doubly charmed baryon was discovered by the SELEX Collaboration~\cite{SELEX:2002wqn}, which motivated further theoretical studies on $cc\bar{q}\bar{q}$ tetraquark states~\cite{Gelman:2002wf,Vijande:2003ki}.
In 2003, $X(3872)$ was discovered by the Belle Collaboration~\cite{Belle:2003nnu}, which opened a new era in  hadron physics. One of the most promising interpretations of $X(3872)$ is a $D\bar{D}^{\ast}$ bound state. This has further stimulated theoretical studies on $D^{(*)}D^{(*)}$ molecules, i.e., $T_{cc}$~\cite{Janc:2004qn,Navarra:2007yw,Vijande:2007rf,Ebert:2007rn,Yang:2009zzp,Molina:2010tx,Li:2012ss,Feng:2013kea,Luo:2017eub,Wang:2017uld,Junnarkar:2018twb,Maiani:2019lpu,Liu:2019stu}. 
In 2017, the LHCb collaboration reported the doubly charmed baryon $\Xi_{cc}$~\cite{LHCb:2017iph}, which allows the $cc$-diquark mass  precisely extracted by the mass of $\Xi_{cc}$ and predicts a doubly charmed compact tetraquark state above the $DD^*$ mass threshold by 8 MeV~\cite{Karliner:2017qjm}. 
 Taking into account the heavy quark symmetry between $\Xi_{cc}$ and $T_{cc}$,  in Refs.~\cite{Mehen:2017nrh,Eichten:2017ffp} a doubly charmed compact tetraquark state above the  $DD^{\ast}$  mass threshold was also  predicted. 
 
   Very recently,  the LHCb Collaboration reported the discovery of a doubly charmed tetraquark state $T_{cc}^+$ with $I(J^{P})=0(1^{+})$, which is found in the  $D^0 D^0\pi^+$ invariant mass spectrum\cite{LHCb:2021auc}. Its binding energy with respect to the  $D^{\ast+}D^{0}$ mass threshold is found to be $\delta=273\pm 61\pm  5^{+11}_{-14}$~keV and  decay width is $\Gamma=410\pm 165\pm  43^{+18}_{-38}$~keV. In the unitarized Breit-Wigner profile  that takes into account the effect of mass threshold, the binding energy and decay width change to $\delta=360\pm 40^{+4}_{-0}$~keV and $\Gamma=48\pm 2^{+0}_{-14}$~keV, respectively~\cite{LHCb:2021vvq}. 
The $T_{cc}$ state could be either a compact tetraquark state  or a hadronic  molecule  of $DD^*$.  Since the $T_{cc}$ mass is below the mass threshold of $D^{\ast+}D^{0}$ by 273~keV, the molecular picture seems more appealing.

In the present work, we revisit the molecular picture for the $T_{cc}^+$ state. In particular, we study its hadronic and radiative decays to check whether one can obtain a decay width in reasonable agreement with the LHCb measurement. This can serve as a highly nontrivial check on  its nature.

\section{Theoretical formalism}

In our previous work~\cite{Liu:2019stu}, we predicted the existence of two doubly charmed  hadronic molecules, $DD^{\ast}$ and $D^{\ast}D^{\ast}$,  with  $I(J^{P})=0(1^{+})$, in the One-Boson-Exchange(OBE) model where the cutoff $\Lambda=$1.01~GeV was fixed by reproducing the binding energy of $X(3872)$,  assumed to be a $\bar{D}^{\ast}D$ bound state. With the cutoff $\Lambda=$1.01~GeV we obtained one $DD^{\ast}$ bound state below the $DD^{\ast}$ mass threshold  by 3~MeV, which is quite close to the mass of $T_{cc}$.  Actually,  the experimental  mass of $T_{cc}$ can be obtained  with a slightly smaller cutoff, i.e., $\Lambda=0.943$~GeV, as shown in Ref.~\cite{Liu:2020nil}.   Therefore, in the OBE model one can regard the newly observed doubly charmed tetraquark state $T_{cc}$ as a $DD^{\ast}$ bound state.  In the following, we focus on its  decay mechanism. 
Since $T_{cc}$  is doubly charmed, it cannot decay into  charmonium states and light mesons and  the allowed strong decay modes should contain open charm mesons.   However, as the $T_{cc}$ mass is below the $DD^*$ threshold, the decay can only proceed via an off-shell $D^*$, which
would heavily suppress its decay width.  Weak and radiative decays are also possible, but they are much smaller in comparison with the strong decay.  

\begin{figure}[ttt]
\begin{center}
\begin{tabular}{ccccc}
\begin{minipage}[t]{0.2\linewidth}
\begin{center}
\begin{overpic}[scale=0.08]{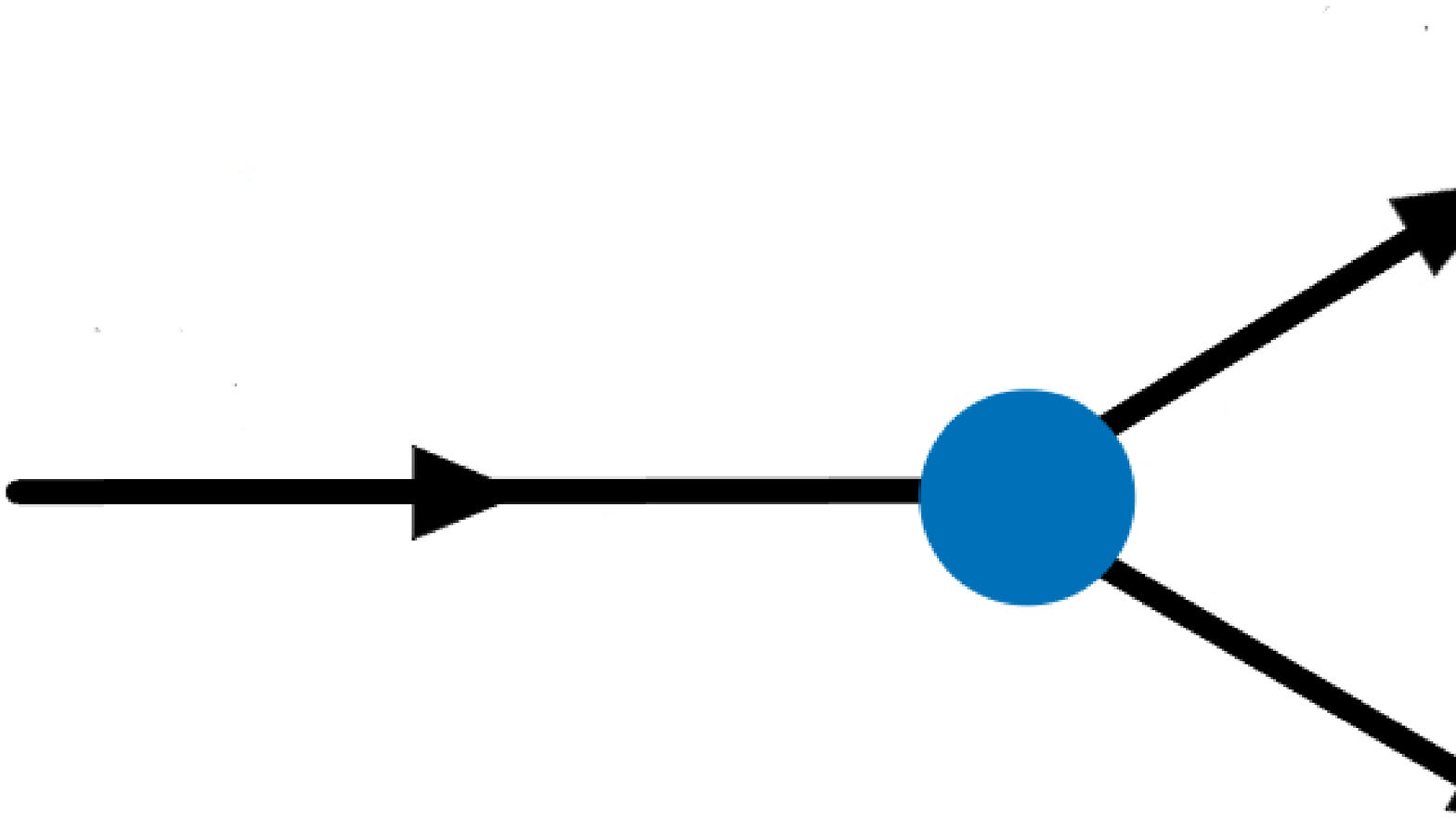}
		\put(10,25){$T_{cc}^+$}
		
		\put(41,3){$D^0$}
		
		\put(41,32){$D^{\ast+}$}
		\put(72,50){$D^+$}
		\put(70,21){$\pi^0$}
				\put(58,-6){(a)}
\end{overpic}
\end{center}
\end{minipage}
&
\begin{minipage}[t]{0.2\linewidth}
\begin{center}
\begin{overpic}[scale=0.08]{tree.eps}
		\put(10,25){$T_{cc}^+$}
		
		\put(41,3){$D^0$}
		
		\put(41,32){$D^{\ast+}$}
		\put(72,50){$D^0$}
		\put(70,21){$\pi^+$}
				\put(58,-6){(b)}
\end{overpic}
\end{center}
\end{minipage}
&
\begin{minipage}[t]{0.2\linewidth}
\begin{center}
\begin{overpic}[scale=0.08]{tree.eps}
		\put(10,25){$T_{cc}^+$}
		
		\put(41,3){$D^+$}
		
		\put(41,32){$D^{\ast0}$}
		\put(72,50){$D^0$}
		\put(70,21){$\pi^0$}
				\put(58,-6){(c)}
\end{overpic}
\end{center}
\end{minipage}
&
\begin{minipage}[t]{0.2\linewidth}
\begin{center}
\begin{overpic}[scale=.08]{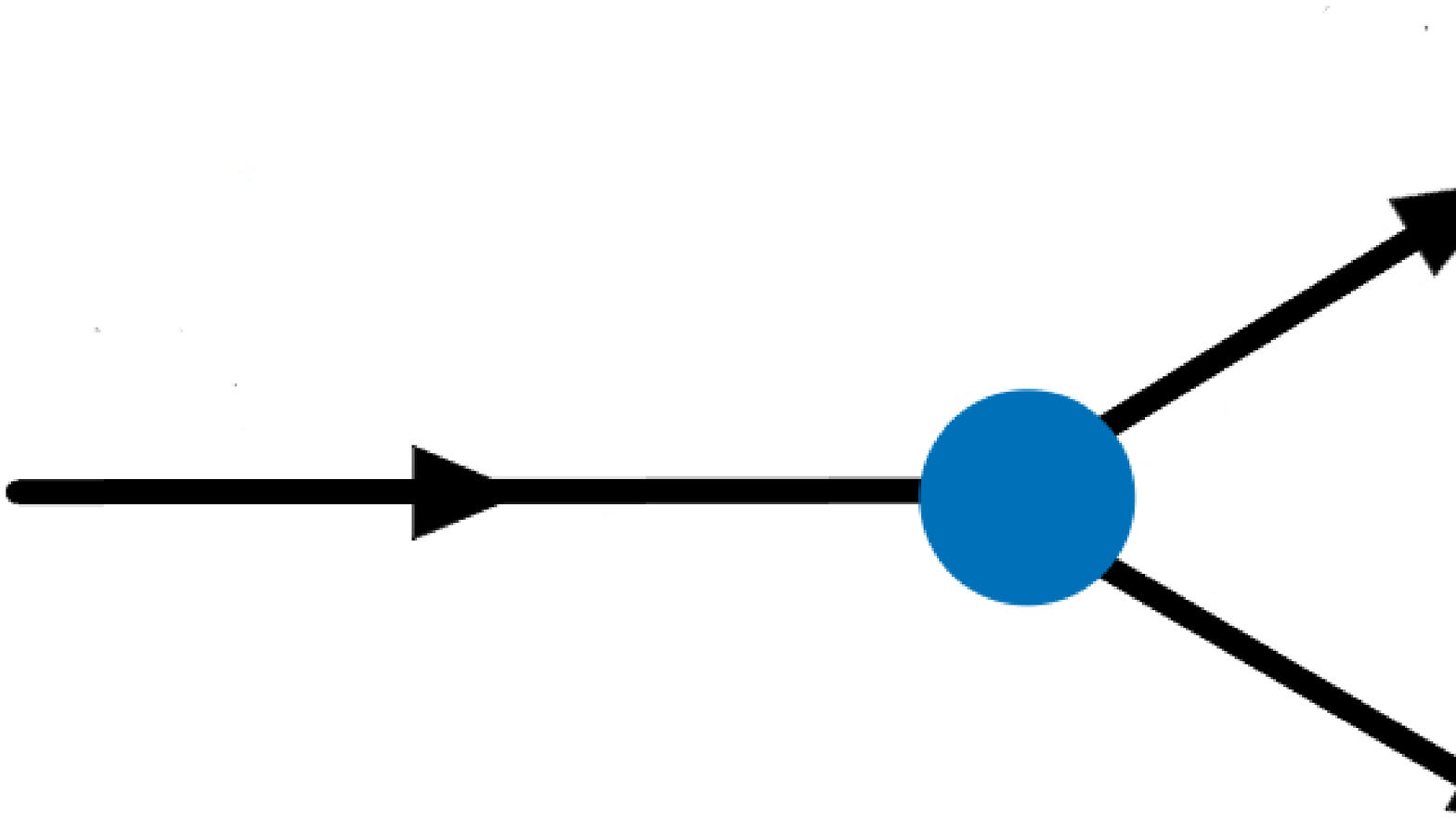}
		\put(10,25){$T_{cc}^+$}
		
		\put(41,3){$D^0$}
		
		\put(41,32){$D^{\ast+}$}
		\put(72,50){$D^+$}
		\put(70,23){$\gamma$}
				\put(58,-6){(d)}
\end{overpic}
\end{center}
\end{minipage}
&
\begin{minipage}[t]{0.2\linewidth}
\begin{center}
\begin{overpic}[scale=.08]{treeg.eps}
		\put(10,25){$T_{cc}^+$}
		
		\put(41,3){$D^+$}
		
		\put(41,32){$D^{\ast0}$}
		\put(72,50){$D^0$}
		\put(70,23){$\gamma$}
		\put(58,-6){(e)}
\end{overpic}
\end{center}
\end{minipage}
\\ \\ 
\end{tabular}
\caption{Tree-level diagrams for strong decays of  $T_{cc}^{+}\to D^{+}\pi^{0}(D^{\ast+})D^{0}$~(a), $T_{cc}^{+}\to D^{0}\pi^{+}(D^{\ast+})D^{0}$~(b)  and  $T_{cc}^{+}\to D^{0}\pi^{0}(D^{\ast0})D^{+}$~(c) as well as  radiative decays of  $T_{cc}^{+}\to D^{+}\gamma(D^{\ast+})D^{0}$~(d) and $T_{cc}^{+}\to D^{0}\gamma(D^{\ast0})D^{+}$~(e).  }
\label{tree}
\end{center}
\end{figure}

In this work, we assume that $T_{cc}^{+}$ is generated by couple channels $D^{\ast+}D^{0}$ and $D^{\ast0}D^{+}$, and it  then can  decay into $D^{+}D^{0}\pi^{0}/D^{0}D^{0}\pi^{+}$ and $D^{+}D^{0}\gamma$ via the  tree-level diagrams, as shown in Fig.~\ref{tree}. In the following, we employ the effective Lagrangian approach to calculate the partial decay widths of $T_{cc}\to D^{+}D^{0}\pi^{0}/D^{0}D^{0}\pi^{+}$ and $D^{+}D^{0}\gamma$.   
 
The interaction between the $T_{cc}$ state and the $DD^{\ast}$ pair is described by the  following effective Lagrangian~\cite{Dong:2008gb}
\begin{eqnarray}
\mathcal{L}_{T_{cc}}(x)=i g_{T_{cc}} T_{cc}^{\mu}(x) \int dy \Phi(y^{2}) D(x+ \omega_{D^{\ast}}y) D^{\ast}_{\mu}(x-\omega_{D}y), 
\end{eqnarray}
where $\omega_{D^{\ast}}=\frac{m_{D^{\ast}}}{m_{D^{\ast}}+m_{D}}$ and $\omega_{D}=\frac{m_{D}}{m_{D^{\ast}}+m_{D}}$ are the kinematic parameters with $m_{D^{\ast}}$ and $m_{D}$  the masses of $D$ and $D^*$, and  $g_{T_{cc}}$ is the coupling between $T_{cc}$ and the ${D}^{\ast}D$ component. The correlation function $\Phi(y^2)$ is introduced to reflect the distribution of the two constituent hadrons in a molecule, which also renders the Feynman diagrams
ultraviolet  finite.  Here we choose the Fourier transformation of the correlation
function in form of a Gaussian function 
\begin{eqnarray}
\Phi(p^2)= e^{-p_{E}^2/\Lambda^2},
\end{eqnarray}
where $\Lambda$ is a size parameter,  which is  expected to be around 1~GeV~\cite{Huang:2019qmw,Ling:2021lla}, and $P_{E}$ is the Euclidean momentum.    The coupling of $g_{T_{cc}}$ can be estimated by reproducing the binding energy of the $T_{cc}$ state via the compositeness condition~\cite{Weinberg:1962hj,Salam:1962ap,Hayashi:1967bjx}.  The condition indicates that  the coupling constant can be determined from the fact that the renormalization constant of the wave function of a composite particle  should be zero. 

For a spin-1 meson, the self energy  can be divided into  a transverse part and a longitudinal part, i.e.,
\begin{eqnarray}
\Sigma^{\mu\nu}=g^{\mu\nu}_{\bot} \Sigma^{T}(k_{0}^{2})+\frac{p^{\mu}p^{\nu}}{p^2}\Sigma^{L}(k_{0}^{2}),
\end{eqnarray}
with $g^{\mu\nu}_{\bot}=g^{\mu\nu}-\frac{p^{\mu}p^{\nu}}{p^2}$.
The compositeness condition  can then be estimated from the transverse part of the self energy
\begin{equation}
   Z_{T_{cc}}=1-\frac{d \Sigma_{T_{cc}}^{T}(k_{0}^{2})}{d{k}_0^{2}}|_{{{k}_0=m_{T_{cc}}}}=0.
\label{23}
\end{equation}

\begin{table}[ttt]
\caption{Masses, quantum numbers and partial decay widths of relevant mesons used in this work~\cite{Zyla:2020zbs}.}
\begin{tabular}{ccc|ccc}
  \hline\hline
   Meson & $I (J^P)$ & M (MeV) &    Meson & $I (J^P)$ & M (MeV)   \\
  \hline
   $D^{0}$ & $\frac{1}{2}(0^-)$ & $1864.84\pm0.05$  &    $D^{+}$ & $\frac{1}{2}(0^-)$ & $1869.66\pm0.05$ \\
  $D^{\ast0}$ & $\frac{1}{2}(1^-)$ & $2006.85\pm0.05$ &  $D^{\ast+}$ & $\frac{1}{2}(1^-)$ & $2010.26\pm0.05$   \\
  $\pi^{+}$ & $1(0^-)$ & $139.57039\pm0.00018$ & 
  $\pi^{0}$ & $1(0^-)$ & $134.9768\pm0.0005$ \\
  \hline 
     Decay mode   & &    Width (keV) &  Decay mode     &&    Width (keV)  \\
 $D^{*+}\to D^0\pi^+${$(D^{+}\pi^{0})$}&& $56.5\pm1.2$($25.6\pm0.6$) & $D^{*+}\to D^+\gamma$&& $1.33\pm 0.03$ \\
  $D^{*0}\to D^0\pi^0$&& $34.658$~\cite{Rosner:2013sha} & $D^{*0}\to D^0\gamma$&& $21.242$~\cite{Rosner:2013sha} \\
 \hline \hline
\end{tabular}
\label{tab:masses}
\end{table}

\begin{figure}[htbp]
\centering
\includegraphics[scale=0.40]{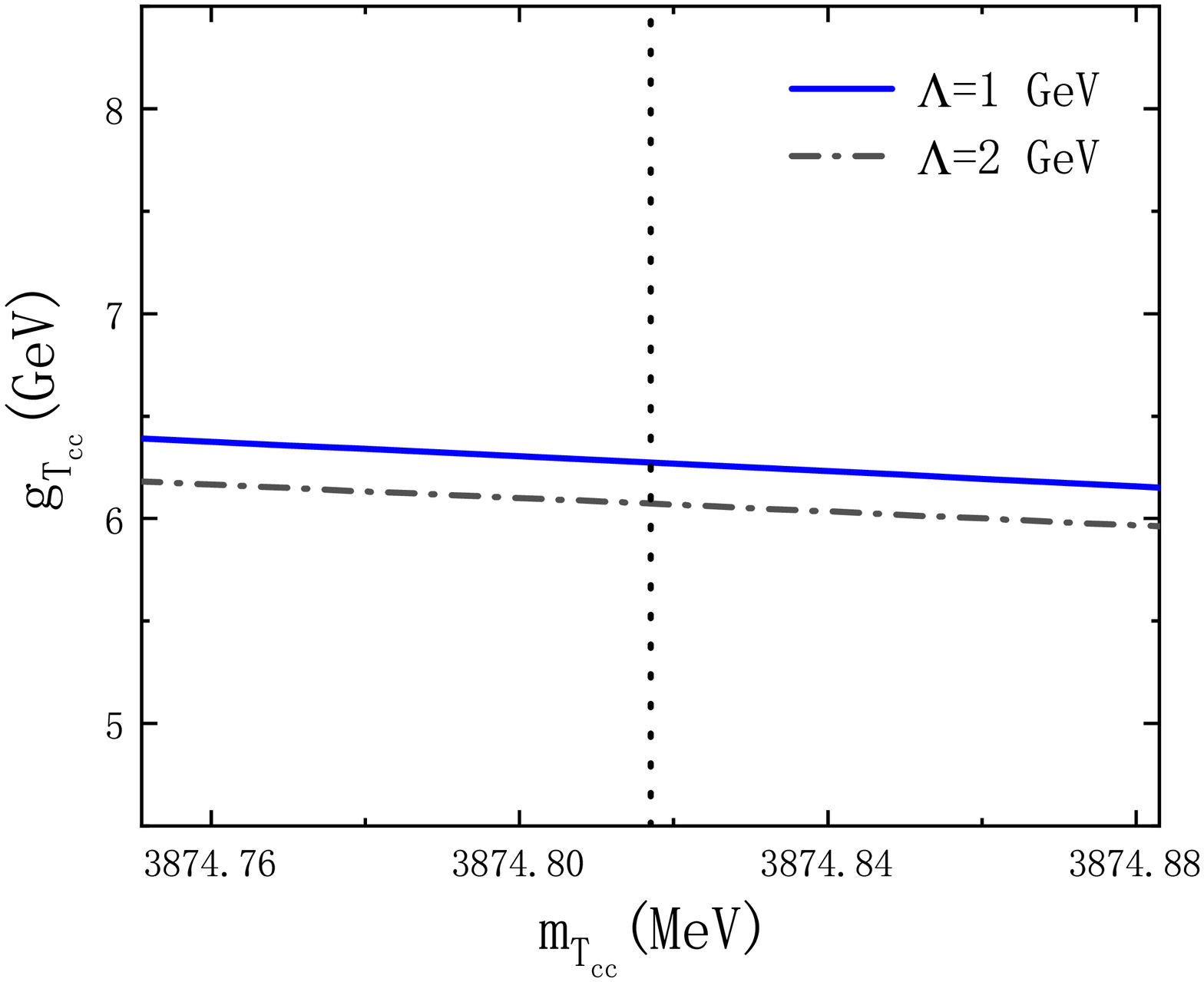}
\caption{Coupling of $T_{cc}$ to $DD^{\ast}$ as a function of the $T_{cc}$ mass  with $\Lambda=1$   and $\Lambda=2$  GeV. The vertical dashed line indicates the experimental central value for the $T_{cc}$ mass.}\label{g1}
\end{figure}

The  $T_{cc}$ mass is below the  $D^{\ast+}D^{0}$ mass threshold by 273 keV with an uncertainty of 66 keV.  We note that the  threshold of $D^{\ast0}D^{+}$  is above that of  $D^{\ast+}D^{0}$ by only about 1.4 MeV, and therefore we assume that the couplings of $g_{T_{cc}D^{\ast0}D^{+}}$ and $g_{T_{cc}D^{\ast+}D^{0}}$ are the same, from  SU(2)-isospin symmetry. Therefore,   we take the average mass of $D^{\ast0}$ and $D^{\ast+}$ ($D^{0}$ and $D^{+}$) to calculate the coupling of $T_{cc}$ to its component $g_{T_{cc}D^{\ast}D}$ (as a result, the binding energy is 0.978 MeV), then obtain  $g_{T_{cc}D^{\ast0}D^{+}}$ and $g_{T_{cc}D^{\ast+}D^{0}}$ using isospin symmetry.    In the isospin symmetric limit,   the $T_{cc}^{+}$ couplings to $D^{\ast+}D^{0}$ and $D^{\ast0}D^{+}$   satisfy the following relationship 
\begin{eqnarray}
g_{T_{cc}^{+}D^{0}D^{\ast+}}=-g_{T_{cc}^{+}D^{\ast0}D^{+}}=\frac{1}{\sqrt{2}}g_{T_{cc}DD^{\ast}}. 
\end{eqnarray}
Substituting the coupling $g_{T_{cc}DD^{\ast}}$  estimated by the compositeness condition, the couplings $g_{T_{cc}^{+}D^{0}D^{\ast+}}$  and $g_{T_{cc}^{+}D^{\ast0}D^{+}}$ are  determined, which turn out to be consistent with the result of  the chiral unitary approach~\cite{Feijoo:2021ppq}.      In Fig.~\ref{g1} we present the dependence of the $T_{cc}$ coupling to  $DD^{\ast}$ on the  $T_{cc}$ mass with the size parameter $\Lambda$ fixed at 1 and 2 GeV. The masses of the involved particles are given in Table~\ref{tab:masses}. One can see that the coupling gradually decreases as the $T_{cc}$ mass increases. Note that the coupling is only weakly dependent on the size parameter $\Lambda$.

The Lagrangian describing  the $D^{\ast}$ decay into $D\pi$ and $D\gamma$ are
\begin{eqnarray}
\mathcal{L}_{D D^{\ast} \pi}&=& -i g_{D D^{\ast} \pi} (D \partial^{\mu}\pi D^{\ast\dag}_{\mu}-D_{\mu}^* \partial^{\mu} \pi  D^{\dag}),  \\ \nonumber
\mathcal{L}_{D D^{\ast} \gamma}&=&e g_{D D^{\ast} \gamma}\varepsilon^{\mu\nu\alpha\beta} \partial_{\mu}A_{\nu}\partial_{\alpha}D^{\ast}_{\beta}D,
\end{eqnarray}
where the fine structure constant $\frac{e^2}{4\pi}=\frac{1}{137}$, and relevant couplings are determined as $g_{ D^{\ast+}D^{0}\pi^{+} }=16.818$ and $g_{D^{\ast+} D^{+}  \gamma}=$0.468 GeV$^{-1}$  by  reproducing the decay widths of $D^{\ast+}\to D^{0} \pi^+$  and
$D^{\ast+}\to D^{+}\gamma $~\cite{Zyla:2020zbs}, respectively.
Experimentally, there exists only  an upper limit $\Gamma<2.1$ MeV for the  $D^{\ast0}$ width.  Thus we turn to the quark model~\cite{Rosner:2013sha}, where the strong  and radiative  decay widths  of $D^{\ast0}$ were estimated to be  $\Gamma_{D^{\ast0}\to D^{0} \pi^{0}}=34.658 $ keV and $\Gamma_{D^{\ast0}\to D^{0} \gamma}=21.242 $ keV.~\footnote{We note that the lattice QCD simulation~\cite{Becirevic:2012pf} gave relatively larger values,  i.e., $\Gamma_{D^{\ast0}\to D^{0} \pi^{0}}=53\pm9$ keV and $\Gamma_{D^{\ast0}\to D^{0} \gamma}=33\pm6 $ keV. From isospin symmetry, we expect that the $D^{*0}$ strong decay width be smaller than the $D^{*+}$ strong decay width because the $D^{*0}\to D^+\pi^-$ decay mode is kinematically forbidden. As a result, we do not use these lattice QCD results.}  With these numbers, we obtain the couplings $g_{ D^{\ast0}D^{0}\pi^{0} }=11.688$ and $g_{D^{\ast0} D^{0}  \gamma}=$1.843 GeV$^{-1}$. It is clear that the strong couplings satisfy approximately isospin symmetry while the electromagnetic couplings do not.

With the above Lagrangians the decay amplitudes of $T_{cc}\to DD\pi$ and $T_{cc} \to DD\gamma$ are
\begin{eqnarray}
\mathcal{M}_{T_{cc}\to DD\pi}&=& ig_{T_{cc}}g_{D D^{\ast} \pi}~p_{2\mu} \frac{-g^{\mu\nu}+k_{1}^{\mu}k_{1}^{\nu}/m_{D^{\ast}}^{2}}{k_{1}^{2}-m_{D^{\ast}}^{2}+im_{D^{\ast}}\Gamma_{m_{D^{\ast}}}}\varepsilon_{\nu}(p_{0}),  \\ 
\mathcal{M}_{T_{cc}\to DD\gamma}&=& ig_{T_{cc}}g_{D D^{\ast} \gamma}\varepsilon^{\mu\nu\alpha\beta} ~p_{2\mu}\varepsilon_{\nu}(p_{2})k_{1\alpha} \frac{-g_{\beta\sigma}+k_{1\beta}k_{1\sigma}/m_{D^{\ast}}^{2}}{k_{1}^{2}-m_{D^{\ast}}^{2}+im_{D^{\ast}}\Gamma_{m_{D^{\ast}}}}\varepsilon^{\sigma}(p_{0}), 
\end{eqnarray}
where $p_{2}$, $k_{1}$, and $p_{0}$ are the momentum of $\pi$($\gamma$), $D^{\ast}$, and $T_{cc}$, respectively. The partial decay widths of $T_{cc}\to DD\pi$ and $T_{cc}\to DD\gamma $ as a function of $m_{12}^2$ and $m_{23}^2$~\cite{Zyla:2020zbs} read:
\begin{equation}
d\Gamma =  \frac{1}{(2 \pi)^{3}}\frac{1}{2J+1} \frac{\overline{|\mathcal{M}|^2}}{32 m_{T_{cc}}^{3}} d m_{12}^{2} d m_{23}^{2},
\end{equation}
with $m_{12}$ the invariant mass of $DD$ and $m_{23}$ the invariant mass of $D\pi$ or $D\gamma$ for the $T_{cc}\to DD\pi$ or $T_{cc}\to DD\gamma $ decay, respectively.

\begin{figure}[htbp]
\centering
\subfigure[]
{
\centering 
\begin{overpic}[scale=.34]{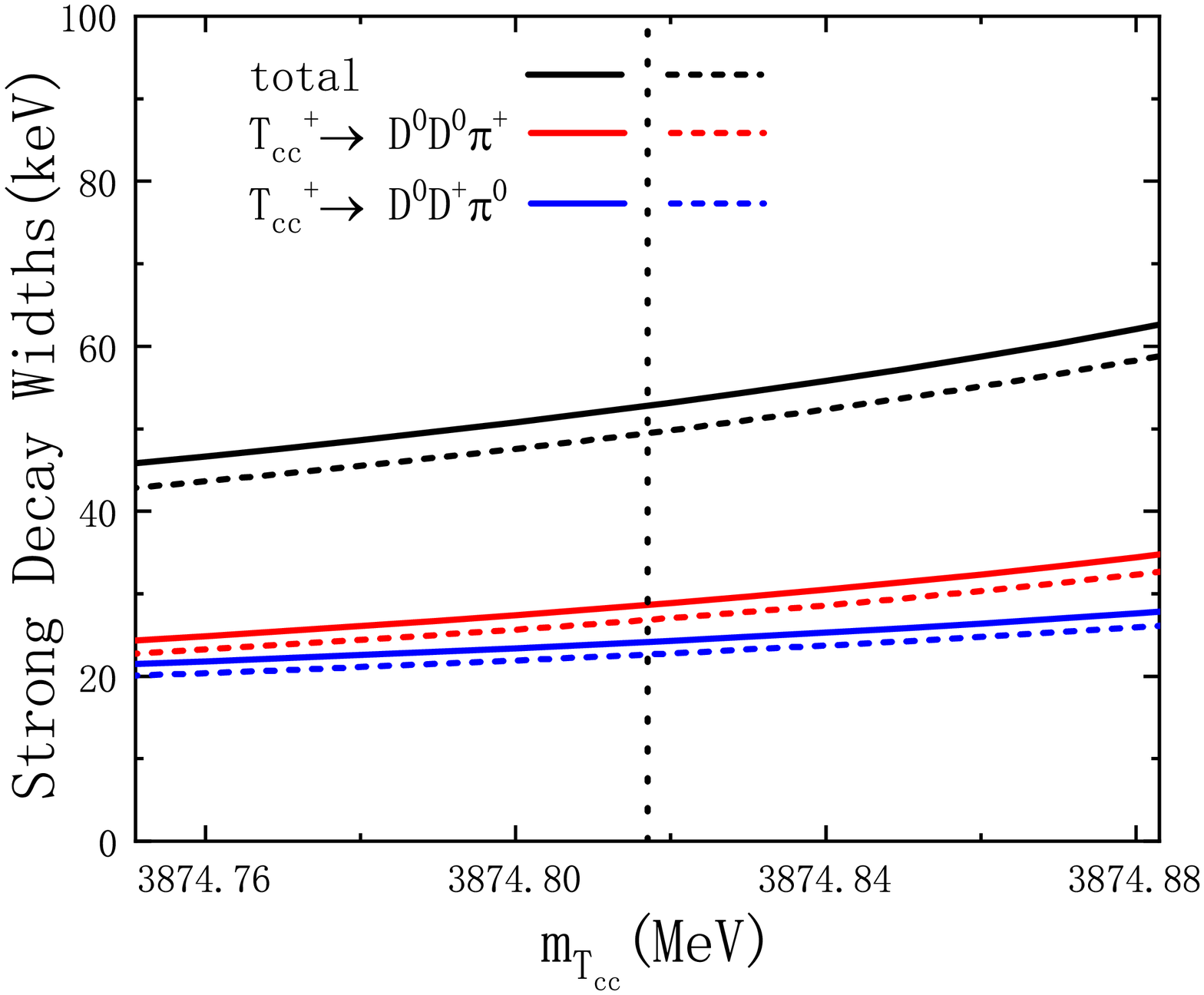}
\end{overpic}
}
\subfigure[]
{
\centering 
\begin{overpic}[scale=.34]{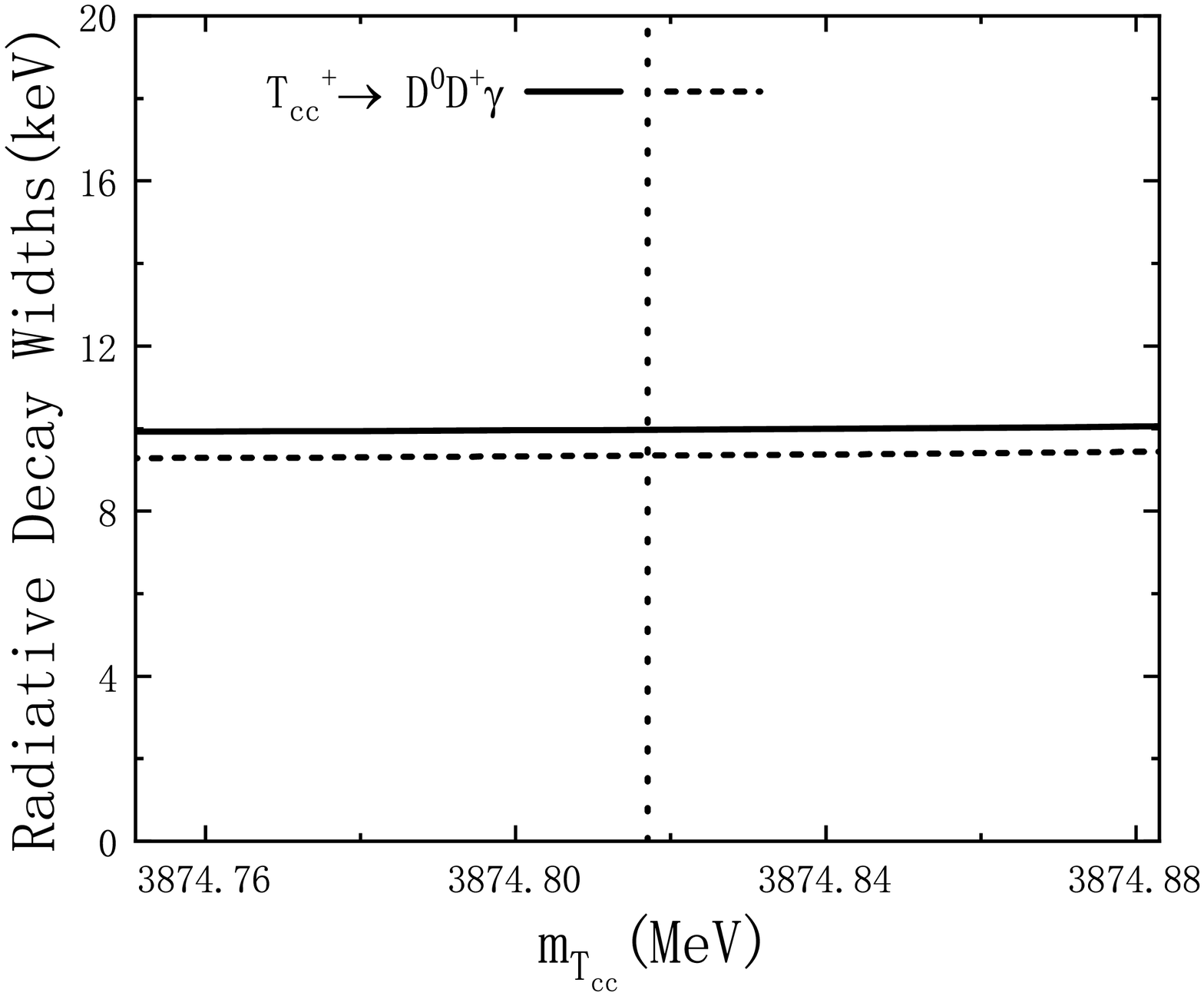}
\end{overpic}
}
  \caption{ Total decay width and partial decay widths  of  $T_{cc}\to DD\pi$(a) and  $T_{cc}\to DD\gamma$ (b) as a function of  the $T_{cc}$ mass. The solid   and dashed lines represent the results obtained with  a cutoff $\Lambda=$1 GeV and $\Lambda=$2  GeV, respectively.  }\label{Gamma}
\end{figure}

In principle, there exist three possible decay channels, $T_{cc}\to D^{0}D^{0}\pi^+$, $T_{cc}\to D^{0}D^{+}\pi^0$, and $T_{cc}\to D^{+}D^{+}\pi^-$. In Fig.~\ref{Gamma}(a/b),  
we show the decay width of $T_{cc}\to DD\pi/\gamma$ as a function of  the $T_{cc}$ mass, where we take the size parameter  $\Lambda=$ 1  and 2 GeV. With the $T_{cc}$ mass varying from 3874.751  to  3874.883 MeV, the   decay width of $T_{cc}\to DD\pi$ is found to be about 46 to 62  keV with the size parameter $\Lambda=1$ GeV,  which is smaller than the Breit-Wigner width  by one order of magnitude~\cite{LHCb:2021auc}, 
but close to the result yielded from the unitary analysis~\cite{LHCb:2021vvq}.  One should note that the decay $T_{cc}(D^{\ast0}D^{+})\to D^{+}\pi^{-}D^{+}$ is kinematically forbidden. The  $T_{cc}(D^{\ast+}D^{0})\to D^{0}\pi^{+}D^{0}$ contribution accounts for 50\% of the total decay width, and the remaining is from $T_{cc}(D^{\ast+}D^{0})\to D^{+}\pi^{0}D^{0}$ and $T_{cc}(D^{\ast0}D^{+})\to D^{0}\pi^{0}D^{+}$.

 The radiative decay width of  $T_{cc}\to DD\gamma$ is about $10$ keV, which mainly originates  from the $D^{*0}D^+$ component of the $DD^*$ molecule  because the radiative decay width of $D^{\ast0}$ is about 16 times of that of $D^{\ast+}$~\cite{Rosner:2013sha}.  Considering isospin breaking due to the mass difference of the $D^{*+}D^0$ and $D^{*0}D^+$ channels, the radiative decay width will decrease. As a result, what we obtained should be viewed as an upper limit.  The ratio of the decay widths of $T_{cc}\to DD\gamma$ and $T_{cc}\to DD\pi$ is in agreement with the ratio of the decay widths of $D^{\ast}\to D \gamma$ and  $D^{\ast}\to D \pi$, reflecting the fact that the decay width of $T_{cc}$ is mainly from the decay of $D^{\ast}$.  Since the decay width of a free $D^{\ast+}$ is 83.4 keV~\cite{Zyla:2020zbs}, the decay width of a weakly bound $DD^*$ molecule is not expected to be larger than 83.4 keV.
 

\section{Summary and outlook}
We studied the decay width of the $T_{cc}^+$
state in the effective Lagrangian approach. Assuming that the $T_{cc}^+$ state is a hadronic molecule of $DD^*$, we obtained a partial decay width of $T_{cc}\to DD\pi$ of about 53 keV and a radiative decay width of about 10 keV.  Their sum is much smaller than the central experimental value of the Breit-Wigner fit, but agrees with that of the unitary fit.   We argued that although the experimental binding energy favors a molecular interpretation for the $T_{cc}$ state, a complete understanding of its decay width is still missing. 

The discovery of  the doubly charmed tetraquark state $T_{cc}^+$ may open up another  new era for hadron physics, in the same way that the discovery of $X(3872)$ did~\cite{Belle:2003nnu}. For open charm exotic baryons, it is very reasonable to expect a complete multiplet of doubly charmed $D^{(\ast)}\Sigma_{c}^{\ast}$ hadronic molecules, which are more bound than their hidden charm $\bar{D}^{(\ast)}\Sigma_{c}^{\ast}$ counterparts~\cite{Liu:2020nil}, consistent with  Refs.~\cite{Dong:2021bvy,Chen:2021kad,Chen:2021htr}.   From SU(3) symmetry, one may expect the existence of $D^{\ast}D_{s}$ or $D^{\ast}D_{s}$ molecules. However, these two systems are found difficult to bind, at least from the perspective of OBE models~\cite{Li:2012ss}.   In a series of recent works~\cite{SanchezSanchez:2017xtl,MartinezTorres:2018zbl,Wu:2019vsy,Wu:2020job}, we predicted one $DDK$  bound state with isospin 1/2, spin-parity  $0^-$, and a minimum quark content of $cc\bar{s}\bar{q}$, which can be regraded as the strangeness partner of $T_{cc}$.    All these remain to be further studied in more detail both theoretically and experimentally. 

  {\it Note added}: After the $T_{cc}^{+}$ state was discovered by the LHCb Collaboration, a  series of works have been performed to investigate the property of $T_{cc}$~\cite{Yan:2021wdl,Fleming:2021wmk,Meng:2021jnw,Chen:2021vhg,Feijoo:2021ppq,Dai:2021wxi,Huang:2021urd,Hu:2021gdg,Albaladejo:2021vln,Jin:2021cxj,Ren:2021dsi,Dong:2021bvy}.   In Refs~\cite{Yan:2021wdl,Fleming:2021wmk}, in addition to the tree level contribution,  the final $DD$ rescattering effect was also taken into account. However, it only contributes  several keV, which obeys the power counting of effective field theory.   In Refs.~\cite{Meng:2021jnw,Chen:2021vhg}, the authors argued that in the molecular picture of $T_{cc}^{+}$ the explicit breaking of isospin should be considered and it may lead to another molecular state mainly coupling to the $D^{\ast0}D^{+}$ channel. However, the authors of Ref.~\cite{Feijoo:2021ppq} did not find another $T_{cc}$ denominated  by the $D^{\ast0}D^{+}$ channel in the chiral unitary model. 
In Ref.~\cite{Dai:2021wxi}, Dai et al. have considered  the interaction of $D^{\ast+}D^{0}$ and $D^{0}D^{0}\pi^{+}$ coupled channels by fitting to the LHCb  data, and interpreted the $T_{cc}^{+}$ state as a virtual state. Moreover, several theoretical works investigated  the production of $T_{cc}$~\cite{Huang:2021urd,Hu:2021gdg}. In Ref.~\cite{Albaladejo:2021vln},   M. Albaladcjo  predicted several $D^{\ast}D^{\ast}$ bound states based on  heavy quark spin  symmetry and   the molecular picture where the $T_{cc}$ state is a $DD^{\ast}$ molecule.

\section{Acknowledgments}
MZL thank Mao-Jun Yan for useful discussions. 
This work is partly supported by the National Natural Science Foundation of China under Grants No.12105007, No.11735003, No.11975041, No.11961141004, No.11961141012, No.12075288, and No.1210050997, and the fundamental Research Funds for the Central Universities, the Youth Innovation Promotion Association CAS, the Key Research Projects of Henan Higher Education Institutions under No. 20A140027, the Project of Youth Backbone Teachers of Colleges and Universities of Henan Province (2020GGJS017),  and the Fundamental Research Cultivation Fund for Young Teachers of Zhengzhou University (JC202041042).
\bibliography{xiccsigmac.bib}

\end{document}